\documentclass{IEEEtran}
\usepackage{hyperref,color,breqn,multirow}%
\usepackage[top=0.7in, left=0.65in]{geometry}
\usepackage{graphicx}
\usepackage{amssymb}
\usepackage{textcomp,gensymb}
\usepackage{comment}
\usepackage[numbers,sort&compress]{natbib}
\usepackage{makecell}
\usepackage{float}
\usepackage{subfigure}
\usepackage{stfloats}
\usepackage{cuted}   
\makeatletter
\renewcommand*{\eqref}[1]{\hyperref[{#1}]{\textup{(\ref*{#1})}}}
\newcommand*{\figref}[1]{\hyperref[{#1}]{\textup{Fig.~\ref*{#1}}}}
\newcommand*{\tabref}[1]{\hyperref[{#1}]{\textup{Table~\ref*{#1}}}}
\newcommand*{\secref}[1]{\hyperref[{#1}]{\textup{Section~\ref*{#1}}}}
\newcommand*{\mat}[1]{\overline{\overline#1}}
\makeatother
\begin{document}

\title{Tunable Perfect Anomalous Reflection Using Passive Aperiodic Gratings}
\author{\IEEEauthorblockN{Yongming Li,~Xikui Ma,~Xuchen~Wang,~Grigorii~Ptitcyn,~Mostafa~Movahediqomi, and Sergei~A.~Tretyakov,~\IEEEmembership{Fellow,~IEEE}}
\thanks{This work was supported by the China Scholarship Council under Grant 202106280229. \textit{(Corresponding author: Xuchen Wang.)}}

\thanks{Yongming Li is with the State Key Laboratory of Electrical Insulation and
Power Equipment, School of Electrical Engineering, 
Xi'an Jiaotong University, Xi'an 710049, China, and with the Department of Electronics
and Nanoengineering, Aalto University, P.O. Box 15500, FI-00076, Espoo, Finland.}

\thanks{Xikui Ma is with the State Key Laboratory of Electrical Insulation and
Power Equipment, School of Electrical Engineering, 
Xi'an Jiaotong University, Xi'an, Shaanxi 710049, China.}

\thanks{Xuchen Wang, Grigorii Ptitcyn, Mostafa Movahediqomi and Sergei A. Tretyakov are with the Department of Electronics
and Nanoengineering, Aalto University, P.O. Box 15500, FI-00076, Espoo, Finland (e-mail: xuchen.wang@kit.edu).}
}


\maketitle

\begin{abstract}
Realizing continuous sweeping of perfect anomalous reflection in a wide angular range has become a technical challenge. 
This challenge cannot be overcome by the conventional aperiodic reflectarrays and periodic metasurfaces or metagratings.  
In this paper, we investigate means to create scanning reflectarrays for the reflection of plane waves coming from any direction into any other direction without any parasitic scattering. The reflection angle can be continuously adjusted by proper tuning of reactive loads of each array element, while the geometrical period is kept constant. 
We conceptually study simple canonical two-dimensional arrays formed by impedance strips above a perfectly reflecting plane. This setup allows fully analytical solutions, which we exploit for understanding the physical nature of parasitic scattering and finding means to overcome fundamental limitations of conventional reflectarray antennas.  
We propose to use subwavelength-spaced arrays and optimize current distribution in $\lambda/2$-sized supercells. As a result, we demonstrate perfect tunable reflection to any angle. 
Our work provides an effective approach to design reconfigurable intelligent surfaces with electrically tunable reflection angles.

\end{abstract}


\thispagestyle{empty}
\pagestyle{empty}

\section{Introduction}
\IEEEPARstart{R}{econfigurable} intelligent surfaces (RISs)  have the prospect to  become one of the key technologies in the next generation of communication systems~\cite{di2020smart, diaz2022integration, wu2019towards}. 
One of the main functionalities of RIS is to redirect an incoming wave to an arbitrary direction, violating the usual reflection law. Such devices are called anomalous reflectors, and they can be realized in several different ways. 
The most well-known method is based on phase-gradient arrays and metasurfaces \cite{Berry_reflectarray,hum2007modeling,Huang2008reflectarray,yu2011light}. In that method, the reflection phase of meta-atoms is designed to vary linearly along the reflector surface, using the locally period approximation (LPA). Unfortunately, the efficiency of reflectors designed using this method drops dramatically as the deflecting angle increases, with strong parasitic reflections especially in the specular direction. In recent years, it has been recognized that the cause of low reflection efficiency is the impedance mismatch between the incident and reflected waves. Perfect anomalous reflection requires optimization of evanescent modes on the surface~\cite{asadchy2016perfect,epstein2016synthesis, kwon2018lossless}. Therefore, more advanced methods for perfect anomalous reflection have been proposed based on non-local metasurfaces~\cite{diaz2017generalized, asadchy2017eliminating, he2022perfect} and metagratings \cite{ra2018reconfigurable, epstein2018metagratings, popov2019beamforming, popov2019constructing, popov2019designing,  casolaro2019dynamic, wang2020independent}.

Most approaches to the design of metasurfaces and metagratings for reflecting plane wave coming from a certain direction to a plane wave traveling in an arbitrary direction are based on periodic structures. In this case, the induced currents and scattered fields can be decomposed into spatial Fourier series (Floquet-Bloch series). Only a finite number of Floquet modes can propagate in space and reach the far zone, and the design goal is to maximize the amplitude of the mode that propagates in the desired direction and minimize the amplitudes of all other propagating harmonics.  The period of the metasurfaces and metagratings is determined by the angle of incidence and angle of reflection. Hence, it is quite difficult to electronically scan the angle of reflection, because scanning requires a continuous change of the array period. However, this difficulty can be overcome by using reflectarrays. The geometrical size of the unit cells of reflectarrays is fixed, usually to $\lambda/2$. Known reflectarrays are designed based on the conventional phased-array principle which has been used since the 60's \cite{Berry_reflectarray, Huang2008reflectarray}. In these devices, the scanning is realized by tuning reactive loads connected to each array element, with the goal to realize a linearly varying phase profile of the induced current along the array plane. The loads are optimized using the local periodic approximation (LPA), numerically simulating the normal-incidence reflection coefficient from a single unit cell in an infinite periodical lattice. Similarly to periodic metasurfaces designed by LPA, the efficiency of reflectarray antennas drops dramatically if the relationship between the incidence and reflection angles deviates much from the usual reflection law. In particular, when the incidence angle is zero, the efficiency becomes low if the desired angle of reflection is greater than approximately  $45^\circ$~\cite{Huang2008reflectarray, Budhu2011understanding, Laurin2013specular}. Beyond this limitation, the parasitic propagating modes carry parts of the energy to undesired directions. As follows from the reciprocity principle, this problem is the same as the well-known problem of parasitic specular reflection in off-set fed reflectarrays. 
 
While for infinitely periodic metasurfaces, solutions to this problem are known~\cite{estakhri2016wave, epstein2016synthesis, asadchy2016perfect, wong2018perfect, kwon2018lossless}, parasitic scattering from reflectarrays is not fully understood (different reasons for this phenomenon are identified by different authors \cite{Laurin2013specular, Budhu2011understanding}), and there are no known means to ensure perfect anomalous reflection to any desired angle using reflectarrays with $\lambda/2$-spaced cells. Here we test a hypothesis that for aperiodic reflectarrays, the optimal loads of usual $\lambda/2$-spaced cells are complex-valued, and seek for solutions to design reflectarrays that can perfectly reflect an incident wave to arbitrary directions by using purely reactive loads. To this end, we investigate analytically a 2D model of a tunable reflectarray, formed by impedance-loaded thin wires or strips above a ground plane. Within this model, for any desired distribution of induced currents in the strips, we can analytically solve for the required load impedance of each element.  Similar structures were studied as periodic planar metagratings without scanning capabilities~\cite{ra2017metagratings, epstein2017unveiling, ra2018reconfigurable, epstein2018metagratings, rabinovich2018analytical, popov2018controlling, rabinovich2019experimental, popov2019beamforming, popov2019constructing, casolaro2019dynamic, xu2021analysis, hansen2022non}, 
but we consider scanning finite-sized arrays with aperiodic loading.
We use these analytical solutions to find a way to design purely passive (lossless) and perfectly operating reflectarrays for an arbitrary reflection angle. In addition to the analytical solution, the scattered electric field distribution is calculated by commercial software (COMSOL multiphysics). 

We show that for a $\lambda/2$-spaced array the direct solution for required load impedances indeed gives complex values, resembling the case of periodical impenetrable impedance boundaries as anomalous reflectors~\cite{asadchy2016perfect}. We consider the effect on performance created by dropping the real part of the calculated loads. It is seen that the performance can go beyond the fundamental limit for local phase-gradient reflectors, but the parasitic scattering is still significant. To overcome this limitation, we next propose the use of several controllable strips in each supercell of $\lambda/2$ size, which allows optimization of the subwavelength current distributions over the array elements. In particular, we study the arrays with three loaded strips per $\lambda/2$, and show that proper optimization of the induced current in a supercell allows us to realize perfect anomalous reflections in any direction.

The remainder of this article is organized as follows. In \secref{sec:section2}, we define the problem and develop the theory. In \secref{sec:section3}, two different examples are presented to validate the theory by numerical simulations. In \secref{sec:section4}, we show how to realize perfect anomalous reflections using purely reactive loads. In \secref{sec:section5} presents the concluding observations.

\section{Principle and methodology} \label{sec:section2}
\begin{figure}
    \centering
    \includegraphics{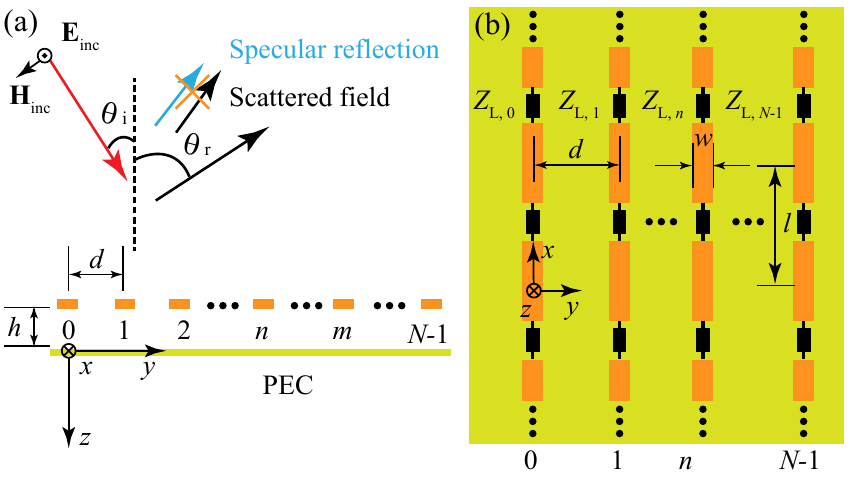}
    \caption{(a) An array of $N$ strips over an infinite ground plane  under a plane-wave illumination at $\theta_{\rm i}$. The desired reflection angle is $\theta_{\rm r}$. The fields radiated by the currents induced in the strips eliminate the specular reflection of the incident wave in the ground plane and launch the desired anomalously reflected plane wave. The distance between strips is $d$. (b) Front view of the array. The strips are loaded by bulk impedances inserted periodically with the period $l$. The width of the strips is $w$.  Both $l$ and $w$ are much smaller than the wavelength in free space. The periodically loaded strips can be modeled as homogeneous impedance strips with the impedances per unit length  $Z_{{\rm L}, n}$, where $n \in \{0, 1, 2, \cdots, N-1\}$.}
    \label{fig:configuration}
\end{figure}

The anomalous reflection problem is illustrated in \figref{fig:configuration}. A finite number of thin conducting strips is placed along the $y$-direction and parallel to the $x$-direction at distance $h$ above an infinite perfect electric conductor (PEC) ground plane. The ground plane is placed at the $z=0$ plane. The strips are equidistantly spaced, and the distance between them (the geometrical period) is $d$. The strips are periodically loaded by bulk reactive loads, and it is assumed that the distance between the load insertions is much smaller than the wavelength of the incident wave. The structure is illuminated by a plane wave at the incidence angle $\theta_{\rm i}$. The incident wave is assumed to be a TE-polarized plane wave with  $\mathbf{E}_{\rm inc}(y,z) = E_0 e^{-j k_0( \sin \theta_{\rm i} y + \cos\theta_{\rm i} z)} \hat{x}$. The time dependence is assumed in the form of $e^{j \omega t}$. In the limit of an infinite number of conducting strips, the goal is to engineer the induced currents in the strips so that the incident wave would be fully reflected to the anomalous direction at  $\theta_{\rm r}$. If we want to generate only one plane wave along the $\theta_{\rm r}$ direction, the simplest solution is to ensure that strips support induced currents whose phase varies linearly with the phase gradient dictated by the reflection angle $\theta_{\rm r}$~(e.g., \cite{hum2007modeling, popov2018controlling}). Indeed, according to the phased-array antenna theory, if the currents flowing in an infinite array of strips have uniform amplitude and a linearly varying phase profile, a plane wave is generated~\cite{tretyakov2003analytical, liu2022reflectarrays}. In addition, it is necessary to eliminate the specular reflection of the incident wave in the PEC ground plane, in order to achieve full-power reflection in the desired direction. For a finite number of strips, such design ensures the highest directivity of the beam reflected towards  $\theta_{\rm r}$ (equivalent to the pattern of a uniformly fed active phased array).

In practice, thin conducting strips are modeled as equivalent round wires with the effective radius $r_{\rm eff} = w/4$~\cite{tretyakov2003analytical}. For a single current line $\mathbf{J}= I \delta(y_0, z_0)\hat{x}$ positioned at $(y_0,z_0)$ and parallel to $x$-axis, the scattered electric field at  point $(y,z)$ can be calculated as~\cite{felsen1994radiation, tretyakov2003analytical}
\begin{equation}
    \mathbf{E}_x = -\frac{k_0 \eta_0 }{4} I H_0^{(2)}(k_0 \sqrt{(y-y_0)^2 + (z-z_0)^2}) \hat{x},   
\end{equation}
where $k_0 = \omega \sqrt{\mu_0 \epsilon_0}$ is the wavenumber in free space, $\eta_0 = \sqrt{\mu_0 / \epsilon_0}$ represents the wave impedance in free space, and $H_0^{(2)}(\cdot)$  is the zeroth-order Hankel function of the second kind. Due to the presence of the ground plane, the electric field produced by strips is created by the induced currents flowing in the strips and their images. The specular reflection of the incident wave from the ground plane produces the field 
\begin{equation}
\mathbf{E}_{\rm ref} = - E_0 e^{-j k_0 (\sin \theta_{\rm i} y-\cos \theta_{\rm i} z)} \hat{x}
. \label{reflected}
\end{equation}
The total external field that excites the strip array is the sum of the incident and the specularly reflected waves, i.e., $\mathbf{E}^{\rm ext} (y,z) = - j 2 E_0 \sin (k_0 \cos \theta_{\rm i} z)e^{-j k_0 \sin \theta_{\rm i} y} \hat{x}$. 
 
In order to eliminate the specular reflection and generate the desired reflected plane wave, two sets of induced currents are required. Let us denote one set of currents that eliminates specular reflection as $\mathbf{J}_\alpha(y, z) = \sum_{m=0}^{N-1}{I_{\alpha} \over d} e^{-j k_0 \sin \theta_{\rm i} y_m}\delta(y - y_m, z + h) \hat{x}$, and the other set of currents that creates the desired anomalously reflected wave as $\mathbf{J}_\beta(y, z) = \sum_{m=0}^{N-1} {I_{\beta} \over d}e^{-j k_0 \sin \theta_{\rm i} y_m}  \delta(y - y_m, z + h) \hat{x}$. Because both current components generate plane waves (in the limit of an infinite array, the complex amplitudes of the strip currents $I_{\alpha,\beta}$ are the same for all strips.  

\subsection{Calculation of  current density distribution}
Let us assume that the current density flowing in strip $m$ is  $\mathbf{J}_m (y,z) = I_m \delta(y-y_m,z + h) \hat{x}$, where $y_m = m d ~(m \in \{0, 1, 2, \cdots, N-1\}$ is the position of the loaded strip. The total current $I_m$ in the strip $m$ has dimension Ampere. The electric field generated by current line $m$ at the position of current line $n$ $(m \ne n)$ is 
\begin{equation}
    \mathbf{E}_{nm} = -\frac{k_0 \eta}{4} I_m \left [ H_0^{(2)} (k_0 y_{mn}) - H_0^{(2)} (k_0 \sqrt{y_{mn}^2 + 4h^2}) \right] \hat{x},
    \label{eq:mutual_electric_field}
\end{equation}
where $y_{mn} = \left| m-n \right| d$ is the distance between the strips $m$ and $n$. The second term of \eqref{eq:mutual_electric_field} is the electric field generated by the mirror image of the strip in the PEC plane. The corresponding mutual impedance between strips $m$ and $n$ can be calculated as $Z_{nm} = \frac{k_0 \eta}{4} \left [ H_0^{(2)} (k_0 y_{mn}) - H_0^{(2)} (k_0 \sqrt{y_{mn}^2 + 4h^2}) \right]$.

According to Ohm's law, the current in the loaded strip with the index $n$ satisfies  
\begin{equation}
    Z_{{\rm L},n} I_n = E^{\rm ext}_x(y_n,-h) - \sum_{m=0,\\m \ne n}^{N-1} Z_{nm} I_m -Z_n I_n,
    \label{eq:Ohm_law}
\end{equation}
where $Z_n={k_0 \eta \over 4} \left[H_0^{(2)}(k_0 r_{\rm eff}) - H_0^{(2)}(2 k_0 h) \right]$ are the self-impedances of strips per unit length. The left-hand side is the total electric field on the strip numbered $n$. The first term on the right-hand side is the external electric field at the strip numbered $n$, \emph{i.e.}, $E_x^{\rm ext} (y_n,-h) =  j 2 E_0 \sin (k_0 \cos \theta_{\rm i} h) e^{- j k_0 \sin\theta_{\rm i} n d}$. The second term on the right-hand side is the field created due to mutual interactions with the other strips. The last term on the right-hand side is caused by its self-action.
The current in each loaded strip has to satisfy \eqref{eq:Ohm_law}, which can be written in a matrix form
\begin{equation}
    \mat{Z} \cdot \vec{I} = \vec{U},
    \label{eq:matrix_Ohm}
\end{equation}
where $\vec{I} = \left[ I_0, I_1, \cdots, I_{N-1} \right]^{\rm T}$ is the vector of induced currents and $\vec{U} =  \left[ E^{\rm ext}_x(y_0,-h), E^{\rm ext}_x(y_1,-h), \cdots, E^{\rm ext}_x(y_{N-1},-h) \right]^{\rm T}$ is the voltage vector. $\mat{Z} = \mat{Z}_{\rm s} + \mat{Z}_{\rm m} +\mat{Z}_{\rm L}$ is the impedance matrix that is composed of  the self-impedance matrix (a diagonal matrix) $\mat{Z}_{\rm s}={\rm diag}\left(Z_0, Z_1, \cdots, Z_n, \cdots, Z_{N-1} \right)$, the load impedance matrix $\mat{Z}_{\rm L} = {\rm diag}(Z_{{\rm L}, 0}, Z_{{\rm L}, 1},\cdots, Z_{{\rm L}, N-1})$, and the mutual impedance matrix $\mat{Z}_{\rm m}$ (see \eqref{eq:Mutual_impedance_matrix}), which has the form
\begin{equation}
    \mat{Z}_{\rm m} =
    \left(
    \begin{array}{ccccc}
         0&     Z_{0,1}&     \cdots&  Z_{0,N-1}\\
         Z_{1,0}      &  0&  \cdots&  Z_{1,N-1}\\ Z_{2,0}      &  Z_{2,1}&     \cdots&  Z_{2,N-1}\\
        \vdots        &  \vdots &        \vdots&    \ddots&   \vdots\\  Z_{N-1,0}     &  Z_{N-1,1}&     \cdots&  0
    \end{array}
    \right).
    \label{eq:Mutual_impedance_matrix}
\end{equation}

As discussed above, the required induced currents flowing in the conductive strips should contain two components, one launches a plane wave to eliminate the specular reflection from the ground plane, and the other generates the desired plane wave. The amplitudes of these two components should be uniform over the whole array. We denote the amplitudes of the first current component as $I_{\alpha}$, and the second component as  $I_{\beta}$. The phases of these two components are set as linear functions along the array plane, to ensure launching waves into the required directions:  $I_m = I_{\alpha} e^{-j k_0 \sin \theta_{\rm i} y_m} + I_{\beta} e^{-j k_0 \sin \theta_{\rm r} y_m}$. Next, we will show how to calculate the required induced currents amplitudes.

\subsubsection{Elimination of the specularly reflected wave} 
To eliminate specular reflection from the ground plane, the sum of the surface-averaged value (the fundamental Floquet harmonic) of the scattered field generated by the induced strip currents and the incident field reflected from the ground should equal zero. We write this condition at $z=-h$, just above the array:  $\mathbf{E}_{\alpha} + \mathbf{E}_{\rm ref} (y,-h) = 0$. 
The electric field generated by the strips can be found using \cite[Eq.~4.35]{tretyakov2003analytical}, which gives 
$
  -\frac{\mathbf{J}_\alpha \eta_0}{2 \cos \theta_{\rm i}}  $. 
The total radiated field is the sum of that electric field generated by the strips and the electric field reflected from the ground:
\begin{equation}
    \mathbf{E}_\alpha = -\frac{\mathbf{J}_\alpha \eta_0}{2 \cos \theta_{\rm i}}  \left( 1 - e^{- j k_0 \cos \theta_{\rm i} 2h} \right).
\end{equation}
Substituting $\mathbf{E}_{\rm ref} (y,-h)$ from \eqref{reflected} and writing $\mathbf{J}_\alpha={I_\alpha\over d}\hat x$, we find 
\begin{equation}
    I_\alpha = j \frac{E_0 d \cos \theta_{\rm i} }{\eta_0 \sin (k_0 \cos \theta_{\rm i} h)}.
    \label{eq:I_alpha}
\end{equation}

\subsubsection{Generation of the desired plane wave}
The averaged scattered electric field generated by the current component $I_\beta$ reads, similarly, 
\begin{equation}
    \mathbf{E}_\beta = -\frac{\mathbf{J}_\beta \eta_0}{2 \cos \theta_{\rm r}} (1 - e^{- j k_0 \cos \theta_{\rm r} 2h}).  
\label{eq:electric_field_beta}
\end{equation}
From  this relation we see that the distance between the plane of strips and the ground plane should satisfy  $h \ne l \frac{\lambda_0}{2\cos \theta_{\rm r}}$, where $l$ is a positive natural number. Otherwise, the sum of the primary field of these currents and the reflection from the ground would equal zero, $\mathbf{E}_{\beta} = 0$, and we could never generate the desired plane wave.

To find the required amplitude of current $\mathbf{J}_\beta$, we will use the power conservation, equating the normal components of the incident plane wave and the anomalously reflected plane wave. First, we calculate the tangential component of the magnetic field that corresponds to the plane wave with the electric field \eqref{eq:electric_field_beta}. This gives  
\begin{equation}
    \mathbf{H}_\beta \cdot \hat y= \frac{E_\beta} {\eta_0 }\cos \theta_{\rm r}.
    \label{eq:magnetic_field}
\end{equation}
The corresponding normal component of the Poynting vector reads $\frac{|E_\beta|^2}{2 \eta_0}\cos\theta_{\rm r}$. Equating that to the normal component of the incident-wave Poynting vector $\frac{|E_{\rm inc}|^2}{2 \eta_0}\cos\theta_{\rm i}$, we find the required current amplitude 
\begin{equation}
    \left| {I_{\beta}} \right| =|J_{\beta} |d = \left| \frac{E_0 d \sqrt{\cos \theta_{\rm i} \cos \theta_{\rm r}}} {\eta_0 \sin (k_0 h \cos \theta_{\rm r})} \right|.
    \label{eq:I_beta}
\end{equation}

Naturally, the phase of current $I_{\beta}$ does not affect the  power carried by the reflected  plane wave. However, it does not mean that this phase has no significance in reflector design and applications.  Actually, if there is no need to engineer a specific reflection phase, the phase of current $I_{\beta}$ can be used as an optimization variable. Later, in \secref{sec:section4}, we will see that when the real parts of load impedances are neglected, the efficiency can be improved by optimizing the phase of $I_{\beta}$.

\subsection{Efficiency}
Once the induced current distribution is known, the electric field generated by the strip array can be calculated as
\begin{align}
   \mathbf{E}^{\rm strips}=&- \frac{k_0 \eta_0 }{4}  \sum_{m=0}^{N-1} 
 I_m \left[ H_0^{(2)} \left(k_0 \sqrt{\left( y - y_m \right)^2+ \left( z+h \right)^2 } \right) \right.\notag\\
 &\left. -  H_0^{(2)}\left(k_0 \sqrt{\left( y - y_m \right)^2+ \left(z-h \right)^2 } \right) \right] \hat{x}. 
 \label{eq:electric_field_strips}
\end{align}
 The scattered electric field is calculated by subtracting the incident electric field from the total electric field. For our finite-sized array, the scattered electric field $\mathbf{E}^{\rm sca}$ is the sum of the field generated by the strips $\mathbf{E}^{\rm strips}$ and the specularly reflected field $\mathbf{E}_{\rm ref}$ \eqref{reflected}, \emph{i.e.}, 
 \begin{equation}\mathbf{E}^{\rm sca} = \mathbf{E}^{\rm strips} + \mathbf{E}_{\rm ref}.
 \label{eq:scat}
 \end{equation}

Usually, the efficiency of anomalous reflection is defined for infinite periodical reflectors as the ratio of the power densities carried by the reflected and incident plane waves or amplitude of the tangential component of the reflected wave and incident wave. For a finite-sized array, it is impossible to generate an ideal plane wave that does not attenuate with the direction of propagation. Since the amplitude of the scattered electric field decreases along the propagation direction, we define the power efficiency of anomalous reflection in terms of far-zone fields as
\begin{equation}
    \zeta = \frac{\vert E^{\rm far} (\theta_{\rm r}, \vec{Z}_{\rm L}) \vert ^2} {\vert E_{\rm refer}^{\rm far} (\theta_{\rm r}) \vert ^2},
    \label{eq:efficiency}
\end{equation}
where $E^{\rm far} (\theta, \vec{Z}_{\rm L})$ is the electric field amplitude generated by strips $E^{\rm strips}$ in the far zone along the direction $\theta_{\rm r}$
for a specific set of strip load impedances which can be written into a vector $\vec{Z}_{\rm L} = \left( Z_{{\rm L}, 0}, Z_{{\rm L}, 1}, \cdots, Z_{{\rm L}, N-1} \right)^{\rm T}$.  $E_{\rm refer}^{\rm far} (\theta_{\rm r})$ is the reference far-zone electric field along the same desired direction created by the ideal current distribution of $I_\alpha$ and $I_\beta$ (with uniform amplitude and linearly varying phase) in Eq.~\eqref{eq:electric_field_strips}.

\section{Analytically calculated required load impedances} \label{sec:section3}
Now that we know what induced currents should flow in the wires in order to anomalously reflect a given plane wave into the desired direction, we can calculate the corresponding load impedances that need to be connected to the strips. This can be done algebraically by solving the system of equations \eqref{eq:matrix_Ohm} for the matrix of impedance loads ${\overline{\overline Z }}_{\rm L}$. Indeed, we know the vector of incident fields $\vec U$, the vector of required currents in the strips $\vec I$, and the impedance matrix of the strip array, and the only unknowns are the load impedances. As a particular example, we assume an incident wave with the frequency $f=10~\text{GHz}$ and the amplitude is $E_0 = 1~{\rm V/m}$, while $h = \lambda / 6$ and the incident angle $\theta_{\rm i} = 0 \degree$. 

The model of a conventional reflectarray corresponds to the case when $d=\lambda/2$ (an array of half-wavelength spaced antenna elements loaded by controllable loads). In this case, the solution for loads that ensure excitation of the desired currents  \eqref{eq:I_alpha} and \eqref{eq:I_beta} is unique. Importantly, we find that the required load impedances are complex-valued. That is, although the array is overall lossless, individual antenna elements should either receive or radiate power. This property is well known for anomalously reflected periodical metasurfaces realized as impedance boundaries whose impedance is defined by the desired plane-wave field distribution \cite{kwon2018lossless, diaz2017generalized}. As to reflectarrays, we have found a discussion of this active-lossy response problem in a classic book \cite{tereshin1980synthesis}. Obviously, the realizations of active and lossy antenna elements in reflectarrays are not practically reasonable. For periodical metasurfaces, the solutions have been found in ensuring the excitation of optimized distributions of evanescent fields. Basically, these evanescent fields form surface waves that effectively carry power from the elements that need to receive power to the elements that need to radiate power, in such a way that the perfect functionality with the purely reactive local response of each array element is achieved. For reflectarrays, this problem has not been even properly recognized, and no solutions are known. 

\begin{figure}[!h]
    \centering
    \includegraphics{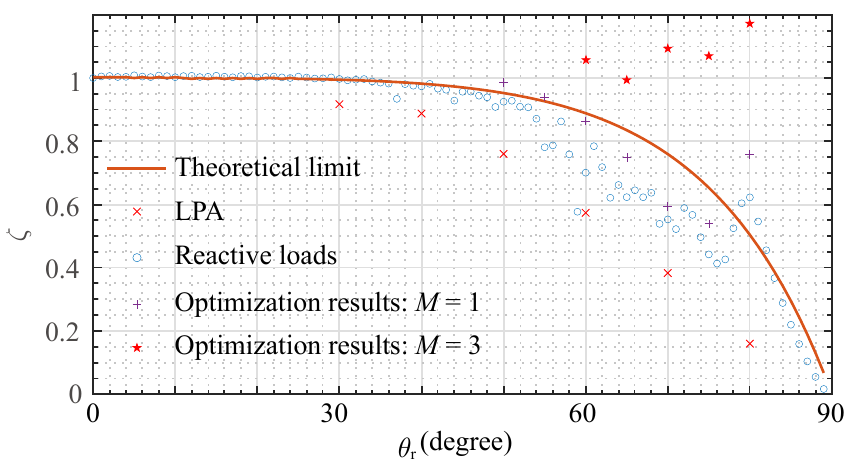}
    \caption{Efficiency as a function of reflection angle $\theta_{\rm r}$. The red line is the theoretical limit for phase-gradient reflectors~\cite{asadchy2017eliminating}. The scatter circles are calculated at a step of $1 \degree$ from $0\degree$ to $89\degree$, when the real parts of the load impedances for $\lambda/2$-spaced array are set to zero. The cross mark '$\times$' is the efficiency of the conventional LPA method. The positive mark '+' represents the optimization results that have only one strip in $\lambda/2$-spaced cells ($M=1$). The red stars are the optimization results which have $3$ strips in $\lambda/2$-spaced cell in \secref{sec:section4} when desired reflection angles are $65\degree$, $70\degree$, and $75\degree$, respectively.}
    \label{fig:efficiency}
\end{figure}

In order to find the solutions for perfect anomalous reflection, first we need to understand how significant is the effect of the required active or lossy response of array elements. We consider the case where the angle of incidence is zero and study the anomalous reflection efficiency as a function of the desired reflection angle.  For the calculated complex-valued loads the efficiency obviously equals unity for all reflection angles, by the construction of the solution. To see the effects of the active-lossy response, we set the real parts of all calculated load impedances to zero and study how the anomalous reflection efficiency changes. The corresponding induced currents are found using \eqref{eq:matrix_Ohm}, and the efficiency is obtained from \eqref{eq:efficiency}. Efficiency $\zeta$ as a function of the reflection angle $\theta_{\rm r}$ is 
 shown in \figref{fig:efficiency} (see blue circles).

As a reference, we plot also the curve showing the fundamental limit for the efficiency of phase-gradient reflectors  defined by \cite{asadchy2017eliminating}
\begin{equation}
    \zeta = \frac{4 \cos \theta_{\rm i} \cos \theta_{\rm r}}{(\cos \theta_{\rm i} + \cos \theta_{\rm r})^2} \label{eq_fund}
\end{equation}
(see red solid line in \figref{fig:efficiency}).  For comparison, we calculate the required load reactances in the conventional locally periodical approximation (LPA) and the efficiency of the corresponding phase-gradient reflector. To do that, we numerically calculate the reflection phase of a periodical array formed by one loaded wire in a $\lambda/2$ cell at normal illumination. Next, we form a phase-gradient reflector loading each wire so that a linearly varying profile is realized in this locally periodical approximation. The efficiencies for several reflection angles are shown by red crosses. It is known that for phase-gradient metasurfaces with subwavelength cells, the efficiency of LPA designs fall below the fundamental limit  \eqref{eq_fund} (a collection of experimental data from the literature can be found in \cite{diaz2017generalized}). We see that this conclusion holds true also for reflectarrays with $\lambda/2$ unit cells. 

Considering the results for analytically calculated load impedances with dropped real parts, we observe that 
for angles $\theta_{\rm r}<30\degree$, the efficiency of anomalous reflection is nearly ideal, and for reflection angles less than $45\degree$, the efficiency remains greater than $95\%$. Thus, this design significantly outperforms phase-gradient reflectors. When the desired reflection angle is small (smaller than about $75\degree$), the efficiency decreases monotonically. For larger reflection angles, efficiency decreases dramatically.  Due to the finite size of the strip array, the efficiency fluctuates up and down when the reflection angle is larger than about $75\degree$, at some point going beyond the fundamental limit for phase-gradient boundaries. 

To better demonstrate the effect of the real part of the loads, let us compare two examples in detail: $\theta_{\rm r} = 30 \degree$ and $\theta_{\rm r} = 70\degree$.

\subsection{\texorpdfstring{$\theta_{\rm r} = 30 \degree$}{}}
\begin{figure*}[!h]
    \centering
    \includegraphics{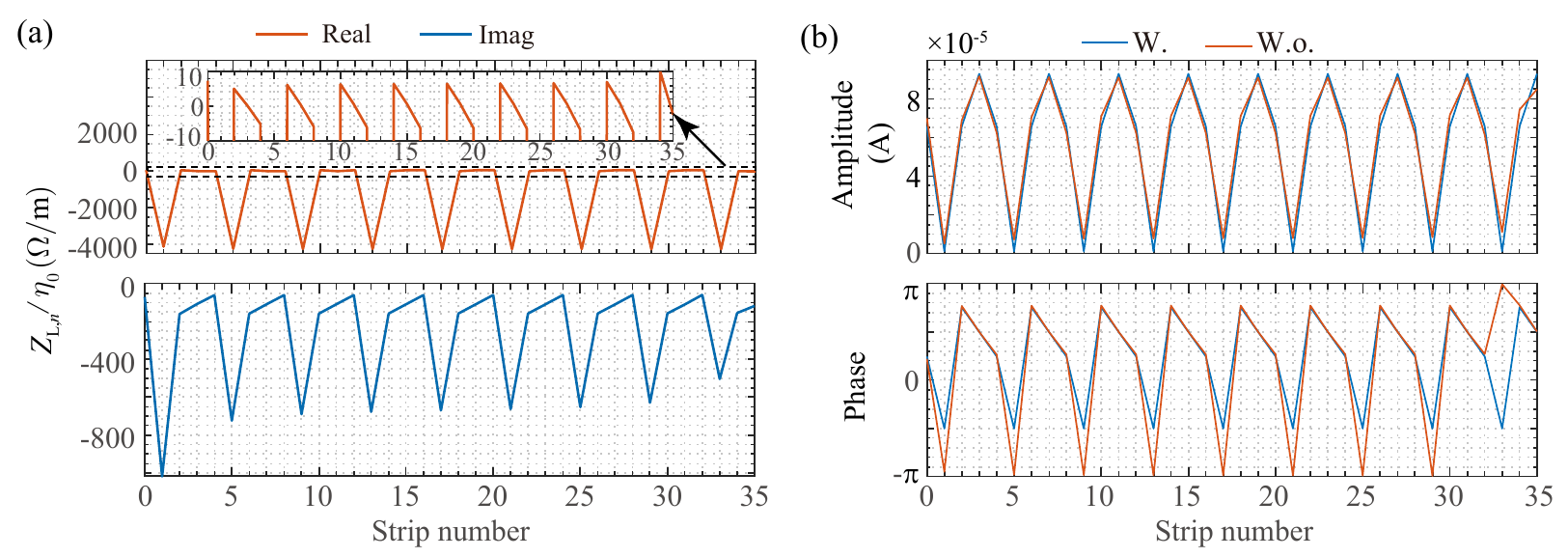}
    \caption{(a) Real and imaginary parts of the load impedance as the function of the strip number when $\theta_{\rm r}=30\degree$. The solid red line corresponds to the real part of the exact complex values of the impedances. The subfigure gives detailed information about the load impedance densities in the black dotted box. The solid blue line corresponds to purely imaginary load impedances. (b) Amplitude and phase of the current distribution with and without the real part of the load impedance densities. The blue line represents the case of complex-valued load impedances, while the red line represents the solution with dropped real parts. We can see that the current has nearly the same distribution both in amplitude and phase.}
    \label{fig:30_degree}
\end{figure*}

In this case, the induced current amplitudes determined by \eqref{eq:I_alpha} and \eqref{eq:I_beta} equal  $I_{\alpha}= j 4.5975 \times 10^{-5} \text{A}$, $\left| I_{\beta} \right| = 4.7045 \times 10^{-5} \text{A}$. In this case, the phase of the induced current is rather special, it has a periodic profile since the chosen period $d=\lambda/2$ matches with the period of periodical reflectors operating at the angle $\theta_{\rm r} = 30 \degree$. The nearly periodic behavior of load impedances and induced currents in the strips can be seen in \figref{fig:30_degree}. It is worth noting that the currents in the strips remain almost unchanged after removing the real part of the loads for such a moderate tilt of the reflected wave (see \figref{fig:30_degree}(b)).

\begin{figure}[!h]
    \centering    \includegraphics{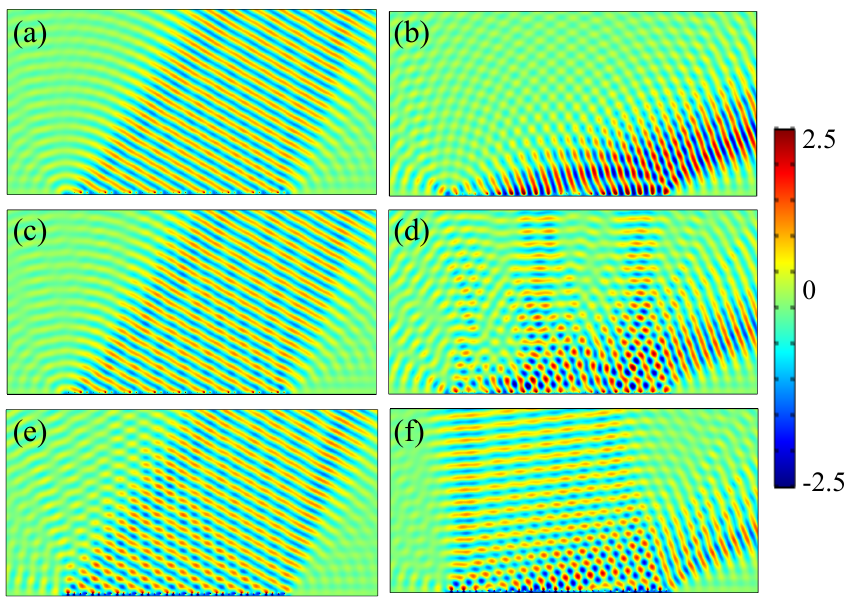}
    \caption{Scattered field distribution $\Re \{E_x^{\rm sca} \}~[{\rm V/m}]$, where (a) and (b) are ideal solutions with active-lossy loads, (c) and (d) show the results when the real part of the loads are dropped,  (e) and (f) are designed with the LPA method.  (a), (c), and (e) are for the case when $\theta_{\rm r} = 30\degree$, (b), (d), and (f) are for the case when $\theta_{\rm r} = 70\degree$, respectively.}    \label{fig:field_distribution}
\end{figure}

Although the induced currents and the corresponding load impedances are calculated analytically, it is not convenient to use the analytical expressions for the scattered fields, because, for finite arrays over an infinite ground plane, the scattered field contains the plane wave reflected from the ground plane outside of the array area. For this reason, for the optimized arrays we compute the fields also numerically, using COMSOL software. In the numerical setup, we simulate the case of a finite-sized ground plane, which allows us to clearly show the distribution of the anomalously reflected fields in space. Details of the simulation setup are given in Appendix~B. 

The real part of the scattered electric field $x$-component distribution is depicted in Figs.~\ref{fig:field_distribution} (a) and (c). It is clearly seen that the scattered fields have nearly the same distribution for exact active-lossy load impedances after dropping the real part of load impedances. This property allows the simple and analytical design of anomalous reflectors based on the known impedance matrix of the array. 
For the conventional LPA design method, it is necessary to calculate the phase of the reflection coefficient for varying loads. In addition, the scattered electric field contains parasitic propagating modes. Using the proposed direct solution for required load impedances, one only needs to use \eqref{eq:matrix_Ohm} and perform simple matrix operations to find the loads and discard the real parts of the obtained values. Compared with the conventional LPA method, this design method is simpler and more efficient (see \figref{fig:efficiency}). 

\subsection{\texorpdfstring{$\theta_{\rm r}=70\degree$}{}}
\begin{figure*}[!h]
    \centering
    \includegraphics{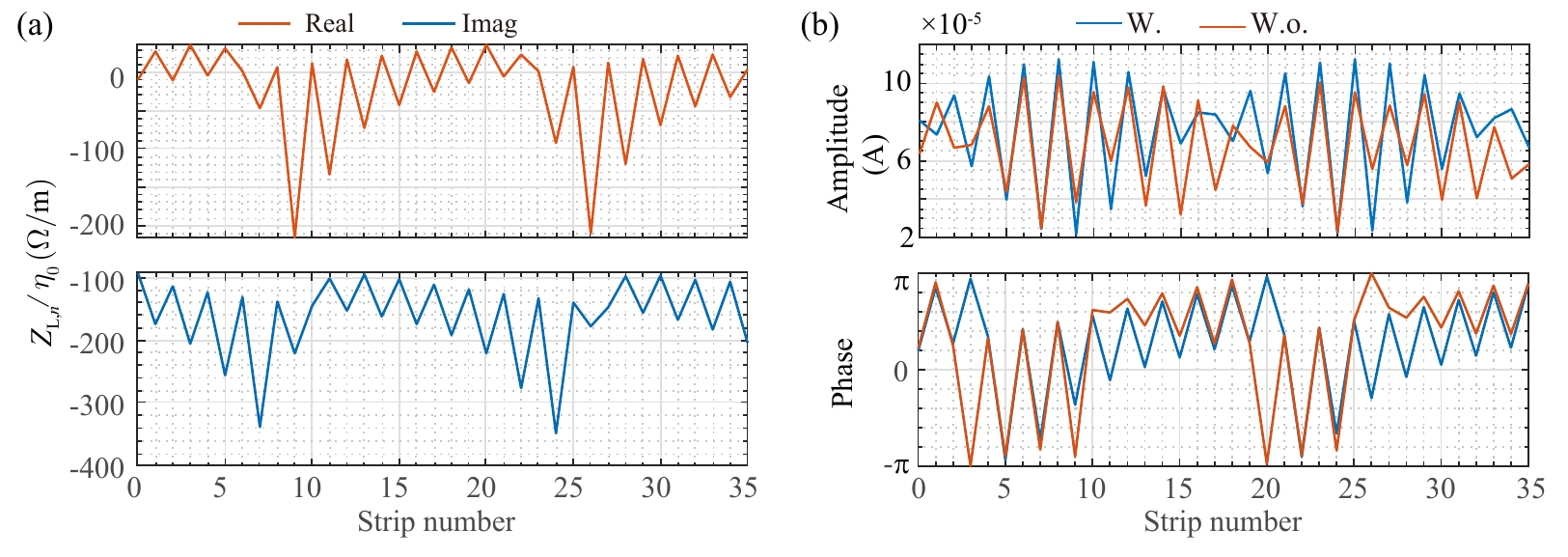}
    \caption{(a) Real and imaginary parts of the load impedance as the function of the strip number when $\theta_{\rm r} = 70\degree$. The solid red line corresponds to the real part of the exact complex values of the impedances,  while the solid blue line corresponds to the purely imaginary load impedance densities. (b) Amplitude and phase of the current distribution with and without the real part of load impedance densities. The blue line represents the case of complex-valued load impedances, while the red line represents the solution with dropped real parts.}
    \label{fig:70_degree}
\end{figure*}

However, it is not always the case. For significantly larger reflection angles, after directly discarding the real part of the loads, the induced currents will be very different, resulting in a drop in efficiency, as part of the energy will be scattered predominantly to the specular direction.

For example, when the desired reflection angle is $\theta_{\rm r} = 70 \degree$, the induced currents determined through \eqref{eq:I_alpha} and \eqref{eq:I_beta} are equal to $I_{\alpha}= j 4.5975 \times 10^{-5} \text{A}$ and $\left| I_{\beta} \right| = 6.6424 \times 10^{-5} \text{A}$.
The load impedance density distribution as a function of the strip number is depicted in \figref{fig:70_degree}(a) while the current distribution is depicted in \figref{fig:70_degree}(b). The induced current distribution is quite different before and after neglecting the real part of the load impedance densities. This obvious change in induced current has a direct impact on the scattered field (see Figs.~\ref{fig:field_distribution}(b) and (d)).

The real part of the scattered electric field $x$-component distribution is depicted in Figs.~\ref{fig:field_distribution}(b) and (d). We can see that when the real part of the load impedance density is ignored, the reflected electric field is significantly distorted and the efficiency drops. Constrained by power conservation, when the real part of the load impedance density is removed, part of the energy is scattered, mostly in the specular direction.

Although this design significantly outperforms conventional phase-gradient reflectors, in this case the efficiency is significantly smaller than unity, and more advanced approaches are needed, which we will introduce next.

\section{Optimized subwavelength arrays to realize perfect anomalous reflectors}
\label{sec:section4}
For large deviations from the usual reflection law (in the considered example of reflection from normal incidence to a tilted angle larger than about $50\degree$), the distribution of the induced current will be quite different before and after neglecting the real part of the required load impedances. The distorted current distribution deteriorates the radiation pattern in the far zone due to the excitation of parasitic propagating modes. In this section, we show that it is possible to dramatically improve the performance of the reflector using optimization of current distribution over unit cells.

As already discussed, for periodical reflectors with the period determined by the desired angles of incidence and reflection, the fundamental limit \eqref{eq_fund} can be overcome by engineering distribution of near-zone (reactive) fields excited by the incident wave \cite{wang2020independent, kwon2021planar}. These known methods, unfortunately, are not applicable for aperiodical, scanning reflectarrays that we consider here. However, it is possible to engineer near-field distributions in Floquet-periodical structures, where the current distribution along the array plane is of the form 
\begin{equation}
    I(y)=F(y)e^{-jk_0\sin\theta\, y}.
    \label{eq_Flo}
\end{equation}
Here, $F(y)$ is a periodical function with the period $\lambda/2$ (or smaller). As is known from the theory of phased-array antennas (e.g., \cite{liu2022reflectarrays}), such current distributions over infinite arrays launch a single plane wave along the direction of the angle $\theta$. By optimizing function $F(y)$, it is possible to control the distribution of evanescent fields in the vicinity of the array, in analogy with the use of subwavelength meta-atoms in periodical metasurfaces, but for scanning arrays with a fixed geometry of the array.

In the frame of our simple model of arrays of thin impedance strips, to gain control over the current distribution in each $\lambda/2$-spaced cell of the array and be able to engineer the evanescent waves, we introduce additional strips in each $\lambda/2$-spaced cell of the structure. It is assumed that $M$ strips are equally spaced in each $\lambda/2$-spaced cell, so that the distance between strips is $\lambda/(2M)$. This approach (with $M>1$) provides the necessary freedom for control of current distribution over unit cells and makes it possible to find loads with purely reactive impedances that can realize perfect anomalous reflection in any direction. The averaged currents in each $\lambda/2$ supercell are set to satisfy   \eqref{eq:I_alpha} and \eqref{eq:I_beta} (with $d=\lambda/2$), so that the specular reflection is properly suppressed and the power conservation in anomalous reflection is satisfied. However, having more than one loaded strip in each unit cell, we can now optimize the subwavelength distribution of the amplitude and phase of the induced current. 

The induced current at the strip numbered $m$ can be expressed in the Floquet-periodic form as
\begin{equation}
    I_m = F_\alpha(y_m) I_{\alpha} e^{-j k_0 \sin \theta_{\rm i} D_m } + F_\beta (y_m) I_{\beta} e^{j \phi} e^{-\text{j} k_0 \sin \theta_{\rm r} D_m}, 
\end{equation}
where $F_\alpha(y_m)$ and $F_\beta(y_m)$ are complex coefficients that define how the total required currents $I_\alpha$ and $I_{\beta}$ in each $\lambda/2$-supercell of the array are distributed between $M$ strips in the supercell. We assume that the anomalously reflected wave has an arbitrary phase shift with respect to the incident field, which we denote as $\phi$. Functions $F_\alpha(y_m)$ and $F_\beta(y_m)$ are periodical with the period $\lambda/2$.  $D_m={\rm floor}(\frac{m}{M})\frac{\lambda}{2}$ defines the linear phase gradient along the array of supercells. The function ${\rm floor}(\cdot)$ returns the smallest integer not greater than its argument. 

The constraints of the optimization are set on the averaged values over supercells, to ensure power conservation. Due to the periodicity of $F(\cdot)$, it is sufficient to give the constraints in one of the $\lambda/2$-sized cells. Here, we give the constraints for the first supercell. The constraints are nonlinear equality constraints and they follow
\begin{subequations}
    \begin{align}
        \sum_{m=0}^{M-1}\Re\{F_\alpha(y_m)\} =& 1\\
        \sum_{m=0}^{M-1}\Im\{F_\alpha(y_m)\} =& 0\\
        \sum_{m=0}^{M-1}\Re\{F_\beta(y_m)\} =& 1\\
        \sum_{m=0}^{M-1}\Im\{F_\beta(y_m)\} =& 0.
    \end{align}\label{eq:ncon}
\end{subequations}
These constraints ensure that the averaged currents in each $\lambda/2$-sized cell have the same magnitude, while the phase of the averaged currents in different $\lambda/2$-sized cells has a linearly varying phase profile, as required for Floquet-periodical current distributions \eqref{eq_Flo}. 

Most importantly, the periodicity of the distribution function $F(y)$ allows us to avoid the need for global optimization of aperiodic arrays. It is enough to optimize the current distribution over only one $\lambda/2$-sized unit cell. Due to the periodicity of $F(y)$, the optimized distribution over one cell determines the required currents induced in all strips of the whole array \eqref{eq_Flo}. Finally, the corresponding load impedances for all strips are found analytically using \eqref{eq:matrix_Ohm}  based on the known impedance matrix of the array.

Thus, instead of more conventional direct optimizations of loads, here we optimize the induced currents in each strip of one unit cell. Optimization is performed by using function \emph{fmincon} in MATLAB. Our goal is to reroute all the incident power in the desired direction. The objective function is to maximize the amplitude of the far-zone electric field in the desired direction. Once the goal is realized, the incident power is rerouted to the desired direction in the most effective way, \emph{i.e.}, perfect anomalous reflection is realized. 

The optimization process is organized as follows: First, an initial set of currents subject to the nonlinear constraints is chosen. Then \eqref{eq:matrix_Ohm} is used to calculate the corresponding load impedances. Since we optimize the induced currents instead of the loads, the obtained loads are in general complex numbers. With the goal to find an optimal solution with purely reactive loads, we 
set the objective function as
\begin{equation}\label{eq:objective function}
O = -{\rm max}\left\{\left| E^{\rm far}(\theta_{\rm r}, \vec{I}(\Im\{\vec{Z}_{\rm L}\}) \right|^2 \right\}.    
\end{equation}
That is, we skip the real parts of the obtained load impedances and calculate the corresponding induced currents. These current values are used to calculate the scattered electric field in the far zone and the objective function. The above steps are repeated until the termination conditions are met.

Let us discuss how many optimization parameters are needed. Because Matlab cannot optimize complex numbers, we count the corresponding real-valued parameters, the real and imaginary parts of the current amplitudes. For periodic structures, the optimization is done in one period, requiring $2$ degrees of freedom for each propagating diffraction order \cite{popov2019constructing}. For the introduced aperiodic realization, the optimization is done in one $\lambda/2$-sized cell. Each set of the induced currents requires $2 M$ parameters since the induced current amplitudes are complex numbers. And we have two sets of induced currents, one for eliminating the specularly reflected wave, and another for generating the desired beam. Hence, the total number of variables to be optimized is $4 M + 1$. An additional variable comes from the phase of current $I_\beta$, which can be used in cases where the application does not require control of the phase of the reflected beam. Taking into account the nonlinear equality constraints, the actual number of independent variables or degrees of freedom is $4 M - 3$. Compared with global optimization, this optimization method has dramatically fewer variables to be optimized, thus making the optimization speed and efficiency much higher, especially when the total number of strips is quite large, for example, $N \ge 10$.

\subsection{Optimized Subwavelength Strip Arrays}
To verify the efficiency of optimization of current distribution inside supercells, we give three examples of optimized anomalous reflectors. We consider the case of $M=3$, that is, in every $\lambda/2$-spaced cell, we position three impedance-loaded strips, so that the distance between the strips is $d=\lambda/6$. There are $36$ supercells, each of which of $\lambda/2$ width, correspondingly, the total width of the array is equal to about $18 \lambda$. The total number of strips equals $108$. In these examples, we use the reflection phase as one of the optimization parameters, thus, we optimize $4M-3=9$ real-valued variables. The array is illuminated by a normally incident plane wave, and the desired reflection angles are $65\degree,~70\degree$, and $75\degree$. 

The optimization is done using the analytical expression for the scattered fields \eqref{eq:scat}, but for the graphical presentation we use numerical simulations, utilizing the same setup as in Section~\ref{sec:section3} (Appendix~B). 

\begin{figure}
    \centering    \includegraphics{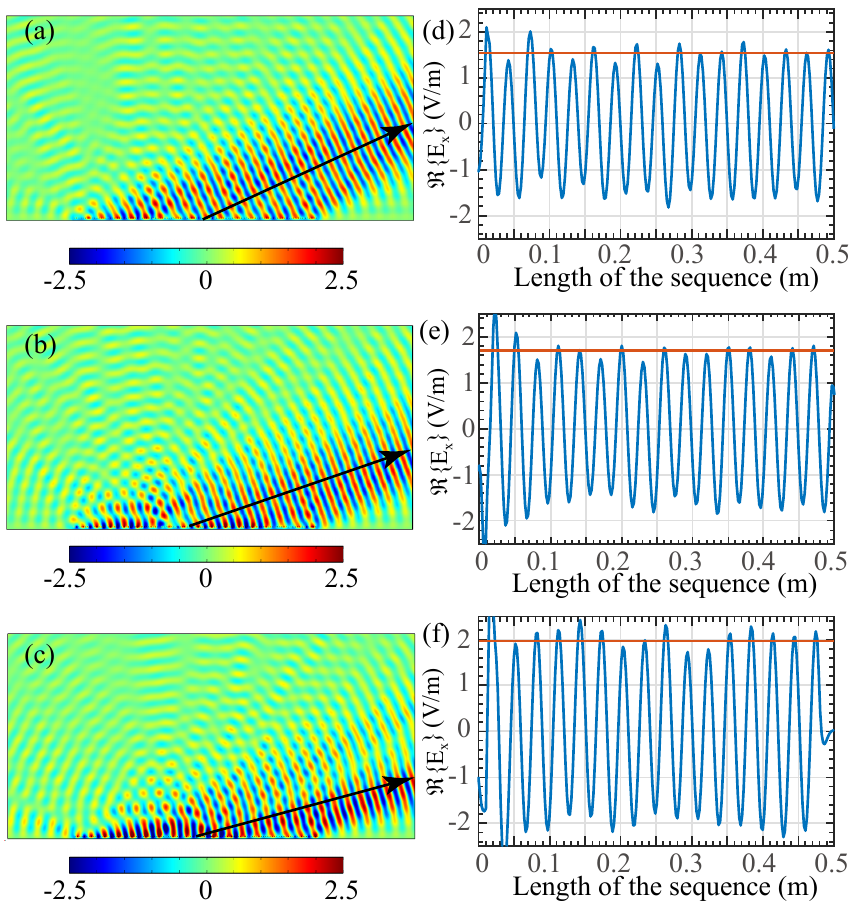}
    \caption{(a) - (c) Real part of the scattered electric field $\Re \{E_x^{\rm sca}\}~[{\rm V/m}]$ distributions after optimization for reflection angles $65\degree,70\degree$, and $75^\degree$, respectively. (d) - (f) Distributions of the real part of the scattered electric field $\Re \{E_x^{\rm sca}\}$ along the black solid arrow shown in subfigures (a), (b), and (c). The red solid lines represent the theoretical optimal amplitudes of the electric field for infinite arrays. }
    \label{fig:optimized_field_distribution}
\end{figure}

The optimized distributions of the scattered electric field $\Re{\{ E_x^{\rm sca} \}}$ for the three considered example reflection directions are shown in Figs.~\ref{fig:optimized_field_distribution} (a), (b), and (c), respectively. In Figs.~\ref{fig:optimized_field_distribution} (d), (e), and (f), we also plot the spatial field variations along the desired reflection direction, shown by the arrows. We compare the field amplitudes with the theoretical values of the required amplitudes of anomalously reflected plane waves from perfect lossless infinite anomalous reflectors. These values are given by  the formula $E_{\rm r} = E_{\rm i} \sqrt{\cos \theta_{\rm i} / \cos \theta_{\rm r}}$ \cite{asadchy2016perfect}, which returns $1.54, 1.71$, and  $1.97$ for  the three considered reflection angles. These values are shown by red lines on the corresponding plots.  The achieved far-field efficiencies, defined by \eqref{eq:efficiency}, equal $99.4 \%$, $109.3\%$, $107.1\%$ for the considered example angles of reflection $65\degree,~70\degree$ and $75\degree$. In these optimizations, the reflection phase $\phi$ is considered as one of the optimization parameters. For the given examples, the phase shifts of the optimized solutions are equal to $\phi=~$0.0\degree$,~$255.5\degree$,~ $43.8\degree$ $ for the angles $65\degree, ~70\degree$, and $75\degree$. 

The results show excellent performance of the arrays of supercells with optimized distributions of induced currents between three elements. The incident power is rerouted to the desired direction, \emph{i.e.}, perfect angle-tunable anomalous reflection is realized using finite-sized arrays with a fixed geometrical period. We also conclude that the optimization leads to weakly super-directive arrays since the efficiency is above unity.  In the last example of reflection to $75\degree$ we see a weak sidelobe at a smaller angle, yet, the far-field efficiency in the desired direction is above unity. Interestingly, it appears that it is possible to realize nearly unit efficiency for all angles even not using the reflection phase as a free variable (fixing the desired phase), but this feature needs more studies.

\subsection{Sparsely Arranged Strip Arrays}
Finally, we compare the performance of the optimized subwavelength supercells with arrays having only one controllable element per $\lambda/2$ cell. In Section~\ref{sec:section3}, we already studied such arrays, analytically calculating the load impedances that realize the desired distribution of induced currents. We see that these impedances are complex-valued, and the performance after skipping the real parts of the load impedances is better than for the conventional phase-gradient reflector. However, this approach to finding the load reactances does not ensure optimal performance, and it is interesting to study if the performance of conventional arrays with a half-wavelength period can be improved by using the proposed optimization approach.

\begin{figure}[!h]
    \centering    \includegraphics{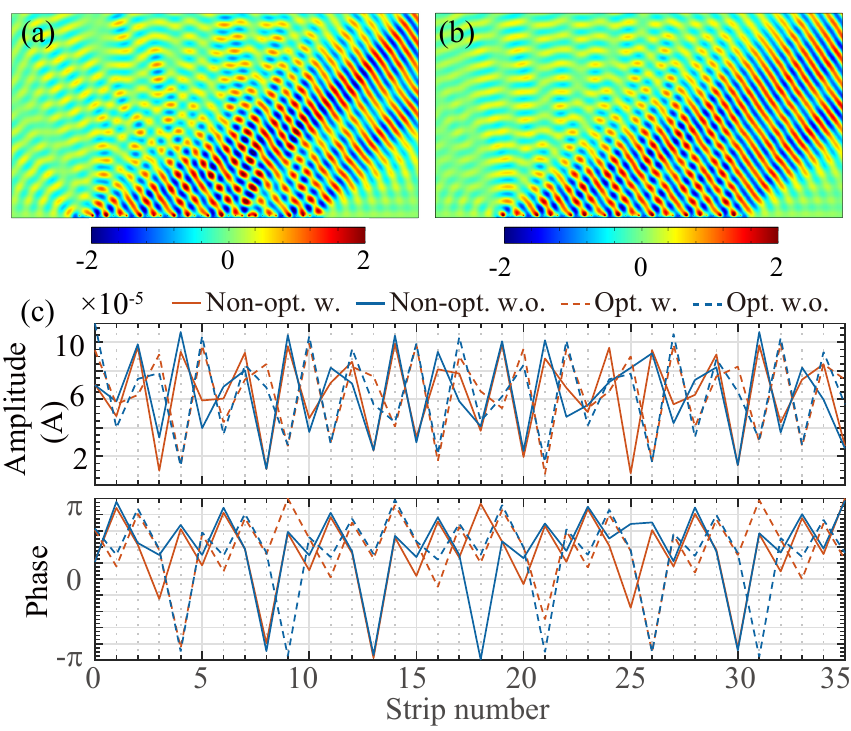}
    \caption{(a) and (b) Scattered electric fields $\Re\{E_x^{\rm sca}\}~[{\rm V/m}]$ for the case $\theta_{r}=55\degree$, when the real part of the loads are neglected. (a) Analytical (non-optimized) solution,  (b) after optimization. (c) Induced current distribution. Red lines show data for complex-valued load impedances and blue curves present the results for reactive loads (real parts dropped).  }
    \label{fig:optimization_M=1}
\end{figure}

For a sparsely distributed strip array $M=1$, \emph{i.e.}, when there is only one strip in $\lambda/2$-spaced cell. This means that only one degree of freedom (the phase of $I_\beta$) can be adjusted to improve reflection efficiency. 
 
The results of reflection efficiency after optimization are shown in \figref{fig:efficiency} with the ’+‘ mark. 
We see that the reflection efficiency achieved with purely reactive loads, in this case, depends on the phase of  $I_\beta$, that is, on the relative phase of the reflected field. 
If there is no need to control this phase, the reflection efficiency can be further improved by properly selecting the phase of $I_\beta$. 

For the case $\theta_{\rm r}=55\degree$, the reflection efficiency is enhanced from $77.3\%$ to $94.0\%$.  It can be seen from Figs.~\ref{fig:optimization_M=1}(a) and (b), that the scattered electric field is more ``clean'' after optimization which suggests a higher efficiency. This result can be explained by considering the corresponding induced current distribution which is depicted in \figref{fig:optimization_M=1}(c). After optimizing the phase of $I_\beta$, the difference between the case with the complex-valued load impedances and the case with the neglected real parts becomes smaller compared with the case without optimization both in the amplitude and phase. Figure~\ref{fig:optimization_M=1}(c) shows that the current distribution after optimization better resembles the ideal one compared to the current distribution obtained by simply dropping the real parts of the load impedances. As a result, the induced currents after optimization provide a higher reflection efficiency. 

We stress that for this case of a single strip in every $\lambda/2$ cell, we cannot realize perfect anomalous reflection. The reason is that we cannot engineer evanescent waves, which are necessary to match the impedances of the incident and reflected waves. As discussed and illustrated above, this problem can be overcome by using densely arranged strip arrays and other means to control the current distribution at the subwavelength scale.

\section{Conclusions} \label{sec:section5}
In summary, we have systematically studied the means to realize finite-sized reflectarrays capable of scanning the reflected beam to any angle without parasitic scattering. The main motivation is the fact that the known solutions for the realization of perfect anomalous reflectors are known only for periodical arrays and metasurfaces that do not allow scanning of the reflection angle. The study is performed using a simple, analytically solvable two-dimensional model of an array of thin strips or wires. First, considering an array of strips with a period $\lambda/2$ (that is, a model of a conventional scanning reflectarray), we calculate analytically the load impedances that need to be connected to each array element in order to realize the desired distribution of induced currents (uniform amplitude and linear phase gradient). The result shows that these loads must be active or lossy. That is, for perfect performance the reflection must be nonlocal. At effectively lossy cells, some part of the incident power is accepted by the array but not re-radiated at the same point. Instead, this power must be transported to another (effectively active) element and be re-radiated there. This is the same phenomenon as is known for the realization of anomalous reflection by using periodic metasurfaces. For scanning reflectarrays with aperiodic distributions of load reactances, this issue was noticed in a classic publication \cite{tereshin1980synthesis}, but the means to overcome this limitation have not been properly explored. It is important to note that this result also explains the physical reason for parasitic specular reflection in off-set reflectarrays. 

One simple possibility that we have considered is to drop the real parts of the required load impedances and use the calculated imaginary parts to set the loads for each element. Naturally, the performance of the array is not anymore perfect, but the results show that it is significantly better than the conventional reflectarrays designed using the local periodic approximation. For reflectors designed to reflect normally incident waves towards arbitrary directions, the reflection efficiency is above $95\%$ for the reflection angles smaller than $45\degree$. The required load reactances are found algebraically, based on the pre-calculated impedance matrix of the array. 

We show also that for a conventional array with $\lambda/2$ cells, the performance of the reflector can be improved by optimizing the phase of the reflected field, but, in that case, it is still impossible to achieve perfect anomalous reflection at extreme angles, because of impedance mismatch between the incident and reflected waves. For periodic metasurfaces, the problem of the required nonlocal response is solved by optimizing surface waves (higher-order Floquet harmonics). Here, we introduce and study the possibility to achieve perfect anomalous reflection by optimizing current distribution over $\lambda/2$-supercells that contain more than one strip.  We have demonstrated that in this way evanescent waves can be engineered for perfect anomalous reflection at any angle. The simulation results have shown the feasibility and effectiveness of this method. Importantly, no global optimization of load reactances is needed, as it is enough to optimize the relative amplitudes and phases of currents inside only one supercell of $\lambda/2$ size. Moreover, the optimization is algebraic after the array impedance matrix is computed (in this work, we used analytical expressions for mutual impedances). In the presented example, perfect performance at all angles, even extreme, is achieved by optimizing only 9 real-valued parameters for a linear aperiodic array containing 108 elements.

Although the subwavelength-scale control is realized by using several subwavelength antenna elements, it is possible to achieve the same goals by using one single patch in each $\lambda/2$ unit cell by introducing several feeding connectors to each patch, so that connecting different reactive loads to different feeds, current distribution over the patch can be controlled. 

Among other benefits, the proposed design method is not constrained to illumination by a single plane wave and reflection towards one desired direction. It can be also used to design reconfigurable beam splitters to achieve different ratios of power distribution between different reflection directions, and other devices for reflection control. Although we demonstrate our method for TE polarized plane waves and scans in one plane only, the physical principles are general, and this approach can be extended for  TM polarization and arbitrary scan directions in both planes. The main target application is in reconfigurable intelligent surfaces for microwave, millimeter, and THz ranges, but the same principles are applicable in optics.

\appendices
\section{LPA method-based anomalous reflection}
We calculate numerically the phase of the reflection coefficient from a periodical array of identically loaded wires as a function of load reactance. Using these results, we form a phase-gradient array loading the wires by reactances needed to realize the desired local reflection coefficient phase. Different loads are assigned to each strip according to different reflection phase requirements at different positions. Next, we use the  relationship between the external electric fields at the wire positions, the load impedances, and induced currents shown in   \eqref{eq:Ohm_law} to calculate the currents induced in all strips from 
\begin{equation}
    \vec{I} = \mat{Z}^{-1} \cdot \vec{U}.
\end{equation}
According to the known induced current distribution, we finally  calculate the scattered electric field and the anomalous reflection efficiency.

\section{Configuration of the aperiodic reflectarray numerical simulation}
\begin{figure}[hb]
    \centering    \includegraphics{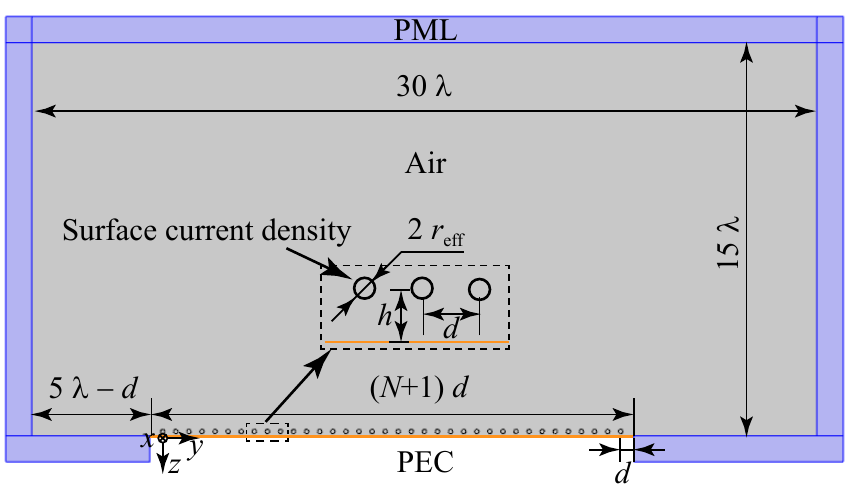}
    \caption{Schematic of the 2D Comsol Multiphysics simulation configuration. The blue area is the perfect matched layer (PML). The surface current density on the wires is defined in terms of an impedance boundary condition, as $\mathbf{J} = E_x/\left(  Z_{{\rm L},n} / (2 \pi r_{\rm eff}) \right) \hat{x}$. }
    \label{fig:COMSOL_configuration}
\end{figure}

The background field $\mathbf{E_{\rm b}}=\mathbf{E_{\rm inc}} = e^{-j k_0 \sin \theta_{\rm i} y - j k_0 \cos \theta_{\rm i} z} \hat{x}$ is used to excite the system. We calculate the corresponding loads using \eqref{eq:matrix_Ohm} and insert these values into commercial software COMSOL Multiphysics to calculate the scattered field. The configuration of COMSOL simulation setting is displayed in \figref{fig:COMSOL_configuration}. 



\bibliographystyle{IEEEtran}
\bibliography{IEEEabrv,Manuscript}

 \end{document}